\documentclass[10pt, letterpaper]{article}
\usepackage[top=1in,left=1in,bottom=1in]{geometry}
\usepackage{authblk}
\usepackage{changepage}
\usepackage[utf8]{inputenc}
\usepackage{textcomp,marvosym}
\usepackage{fixltx2e}
\usepackage{amsmath,amssymb}
\usepackage{cite}
\usepackage{nameref}
\usepackage[pdfpagemode=UseOutlines,bookmarks=T,pdfstartview=FitH]{hyperref}
\usepackage[right]{lineno}
\usepackage{microtype}
\DisableLigatures[f]{encoding = *, family = * }
\usepackage{rotating}
\usepackage{lineno,setspace}

\raggedright
\date{}

\usepackage{lastpage,fancyhdr,graphicx}
\usepackage{epstopdf}
\def\tableline{ \vskip .1in \hrule height .6pt \vskip 0.1in}
\newcommand{\ppr}{\lambda_1(\mathbf{A})}
\newcommand{\ID}{I/D}

\let\proglang=\textsf
\newcommand{\R}{\proglang{R}}  

\author[a,b,c,$\dag$]{Stuart R. Borrett}
\author[a]{Montgomery Carter}
\author[a]{David E. Hines}

\affil[a]{{\small Department of Biology \& Marine Biology, University
  of North Carolina Wilmington, Wilmington,
  NC, 28403 USA}}
\affil[b]{{\small Center for Marine Science, University of North
    Carolina Wilmington, Wilmington, NC}}
\affil[c]{{\small Duke Network Analysis Center, Social Science Research Institute, Duke University, Durham, NC, USA}}

\affil[$\dag$] {{\small Corresponding Author: borretts@uncw.edu}}

\title{Six General Ecosystem Properties are more Intense in
   Biogeochemical Cycling Networks than Food Webs}

\begin{document} 
\maketitle

\tableline
\section*{Abstract}

Network analysis has revealed several whole-network properties
hypothesized to be general characteristics of ecosystems. These
include pathway proliferation, and network non-locality,
homogenization, amplification, mutualism, and synergism.  Collectively
these properties characterize the impact of indirect interactions
among ecosystem elements.  While ecosystem networks generally trace a
thermodynamically conserved unit through the system, there appear to
be several model classes.  For example, trophic (TRO) networks are
built around a food web, usually follow energy or carbon, and are the
most abundant models in the literature.  Biogeochemical cycling (BGC)
networks trace nutrients like nitrogen or phosphorus and tend to have
more aggregated nodes, less dissipation, and more recycling than
TRO. We tested (1) the hypothesized generality of the properties in
BGC networks and (2) that they tend to be more strongly expressed in
BGC networks than in the TRO networks due to increased recycling. We
compared the properties in 22 biogeochemical and 57 trophic ecosystem
networks from the literature using enaR.  We also evaluated the
robustness of these results with an uncertainty analysis.  The results
generally support the hypotheses. First, five of the properties
occurred in varying degrees in all 22 BGC models, while network
mutualism occurred in 86\% of the models. Further, these results were
generally robust to a $\pm$50\% uncertainty in the model parameters.
Second, the average network statistics for the six properties were
statistically significantly greater in the BGC models than the TRO
models. These results (1) confirm the general presence of these
properties in ecosystem networks, (2) highlight the significance of
different model types in determining property intensities, (3)
reinforce the importance of recycling, and (4) provide a set of
indicator benchmarks for future systems comparisons.  Further, this
work highlights how indirect effects distributed by network
connectivity can transform whole-ecosystem functioning, and adds to
the growing domain of network ecology.
\vskip 1em
 \textbf{Keywords:}  network environ analysis; ecological network
 analysis; indirect effects; input--output analysis; systems ecology;
 network science

\tableline
%


\newpage
\begin{spacing}{1}

\section{Introduction}

Complex systems are comprised of reticulated exchanges of energy,
matter, and information that let members influence each other directly
and indirectly.  These connections can define member characteristics
\cite{newman01scientific1, csermely2013, luczkovich03, ulanowicz2014}
and create whole-system properties such as stability \cite{otto2007,
  thebault2010} and tolerance to failure \cite{albert00, dunne02}.
The degree of connectivity is important, but patterns of connectivity
are critical for these effects \cite{barabasi02, dunne02, estrada07,
  scotti2013social}.  Network models are ideal for mapping these
exchanges, and network analyses characterize the subsequent
relationships and quantify the connectivity patterns \cite{newman2010,
  brandes05, barabasi12}.

Ecosystems can be modeled as a network in which nodes represent
species, functional groups of species, or non-living components, and
directed edges trace the direct transactions of energy or matter among
the nodes (Fig.~\ref{fig:example}).  Thus, this network model is like
a road map for the physical exchange of resources.  Ecologists then
analyze the network model to understand the consequences of
whole-ecosystem organization and the often hidden relationships among
species and the external environment \cite{ulanowicz2014, higashi89,
  fath98, bondavalli99, hines15, heymans2014global, saintbeat2015}.
For example, Grami et al. \cite{grami2011} found that in a food web
model of Lake Pavin the fungal parasites of phytoplankton (chytrids)
generally reduced the carbon lost from the pelagic zone.  This effect
was due in part because the chytrids increased the average path length
that the carbon molecules traveled in the system, which they further
hypothesized created greater system stability.  Small et
al. \cite{small2014} used ecosystem networks to compare nitrogen
cycling in the Laurentian Great Lakes, and discovered that Lake
Superior had a higher denitrification efficiency than Lake Huran or
Lake Erie despite its having a low areal denitrification rate.  This
ecosystem network analysis (ENA) builds our understanding of the
importance of connectivity patterns in ecosystems, contributes to the
rise of network ecology \cite{borrett14_rise, bascompte07}, and more
broadly contributes to the emerging domain of network science and its
approach to studying complex systems in general \cite{newman2010,
  nrc06network, barabasi12, brandes13}.

While ecosystem networks are formed by tracing a
thermodynamically conservative unit through the system
\cite{fath07_netconstruction}, there appear to be several classes of
ecosystem network models with distinctive features \cite{christian96,
  baird08_sylt, borrett10_idd}.  
Trophic networks (TRO) are the most common class of ecosystem
networks.  They are built around a food web, and typically trace either
energy or carbon.  The South Carolina oyster reef model \cite{dame81}
and Chesapeake Bay ecosystem models \cite{baird89} are typical
examples of this class.  A second class of ecosystem network model
focuses on biogeochemical cycling (BGC), and typically traces a key
nutrient like nitrogen or phosphorus.  Christian et
al. \cite{christian96} suggested that by comparison to the trophic
models, biogeochemical cycling models tend to have more aggregated
nodes, less dissipation and more recycling.  Models of nitrogen
cycling in the Laurentian Great Lakes \cite{small2014}, Neuse River
Estuary \cite{christian03}, and the Cape Fear River Estuary
\cite{hines12, hines15} are examples of this class.  In this work, we
investigate the generality of selected ecosystem network properties in
the two most abundant classes (TRO and BGC).

\subsection{Six Hypothesized Characteristics of Ecosystem Networks}
Through the development and application of 
ENA, several hypotheses regarding the organization of whole ecosystems
have emerged \cite{ulanowicz86, fath99_review, jorgensen07_newecology,
  jorgensen2012systems}.  Here, we focus on six of these hypotheses:
pathway proliferation, network non-locality, network homogenization,
network amplification, network mutualism, and network synergism.

\begin{figure}[t]
\center
\includegraphics[scale=0.75]{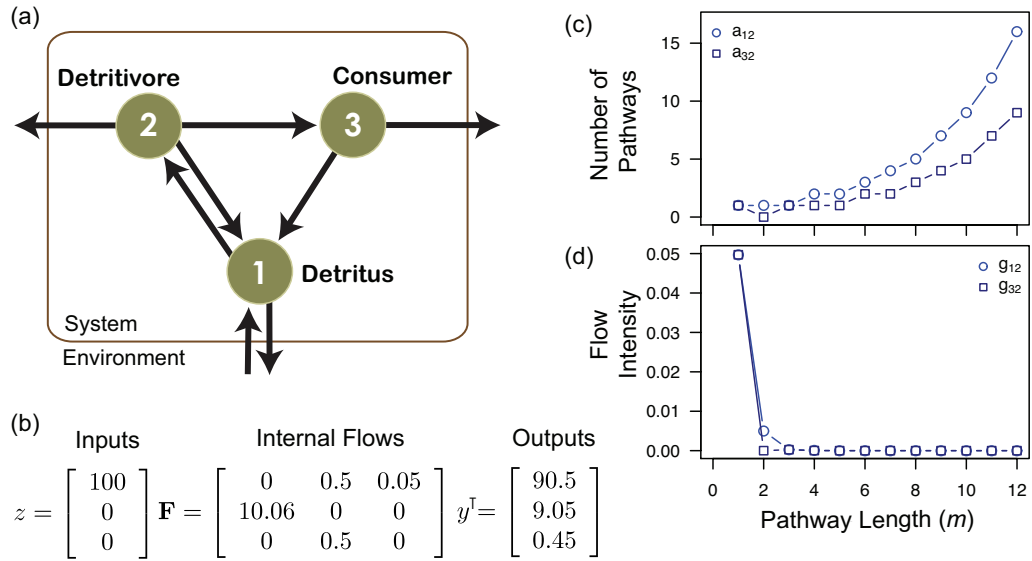}
  \caption{{\bf Hypothetical ecosystem network model}.  (a) The
    directed graph represents an ecosystem with the three nodes
    (Detritus, Detritivores, and Consumer) along with the internal
    flows $f_{ij}$ and boundary inputs $z_i$ and outputs $y_j$ of
    energy or matter.  The matrices (b) summarize the steady-state
    dynamics of energy or matter fluxes [M L\textsuperscript{-2}
    T\textsuperscript{-1}] in the system if we assume a 10\% trophic
    transfer efficiency. Pathway proliferation (c) is shown as the
    number of pathways (walks) from node 2 to 1 ($a_{12}$) and from
    node 3 to 2 ($a_{32}$)
    increase with path length, and (d) shows the expected rapid
    decline in flow intensity ($g_{ij}$) between nodes as path length
    increases. } \label{fig:example}
\end{figure}

\emph{Pathway proliferation} occurs when the number of
pathways in an ecosystem tends to rapidly increase as path length
increases \cite{patten85, borrett03, borrett07_jtb}.  In this
literature a pathway is defined as a sequence of nodes and edges (e.g.,
Detritus $ \rightarrow $ Detritivore $ \rightarrow $ Consumer in
Fig.~\ref{fig:example}a), and the length of the pathway is the number
of edges $m$, where $m = 2$ in the example.  This is sometimes called a
walk in graph theory as both nodes and edges can be repeated
\cite{brandes2005_fun}.  These longer pathways create a reticulated
web of possible routes of influence.  Pathway proliferation is
illustrated in Fig.~\ref{fig:example}c.

\emph{Network non-locality} is the tendency for indirect flows (flows
over pathways where $m>1$) to cumulatively exceed direct flows in
ecosystems \cite{patten83, higashi89, salas11_did}.  This occurs
despite the diminishing flow intensity as path length increases
(Fig.~\ref{fig:example}d) due to the thermodynamically open nature of
ecosystems. The implications of this property are that organisms not
only have the ability to influence each other indirectly, but that
these indirect relations may be both quantitatively and qualitatively
dominant.  Further research has shown that these indirect interactions
have the power to transform the nature of the relationships among
species.  For example, Bondavalli and Ulanowicz \cite{bondavalli99}
found that due to indirect interactions across the trophic network in
the Everglades the American alligator was more beneficial to some of
its prey like frogs than they were detrimental.  This complicates
environmental management and restoration \cite{patten82_EPA,
  bondavalli99}.

\emph{Network homogenization} is the tendency of the indirect flows to
more evenly distribute resources in the system.  Patten et
al. \cite{patten90} observed that when they applied ENA flow analysis,
the integral (boundary + direct + indirect) flow intensities were more
evenly distributed than the direct flow intensities alone.  They
inferred that the indirect flow intensities must be spreading the
resource exchange.

\emph{Network amplification} is defined as ``the increase of
energy--matter ultimately utilized at destination compartments
compared to the quantities originally introduced at source
compartments'' \cite{higashi93modes}.  Though it may at first seem
counter intuitive, this property is made possible by energy and matter
recycling \cite{patten85_energy,higashi93modes}. 
It implies that some exchanges are quite important for the system
function.

\emph{Network mutualism} and \emph{Network synergism} both reflect the
tendency for indirect interactions to make the relationships among the
network nodes more positive than they appear from the direct
interactions \cite{patten91, fath98, fath07_mut}, as with the
previously mentioned case of the frog and alligator.  Network
mutualism examines the number of positive and negative interactions,
while network synergism compares the magnitudes of the positive and
negative interactions.

\subsection{Existing Evidence}
These hypothesized general ecosystem properties were first identified
in relatively small, well-connected, trophic-based ecosystem models
(e.g., oyster reef model \cite{dame81}), but subsequent research has
shown that four of the six properties appear to be general in trophic
models. For example, pathway proliferation is known to occur in all
well connected models \cite{borrett07_jtb}. Higashi and Patten
\cite{higashi86, higashi89} showed algebraically why network
non-locality is likely to occur in ecosystem networks. Fath and Patten
\cite{fath99_homo} introduced a statistic to quantify network
homogenization and showed that the property occurred in two ecosystem
models and a set of arbitrary networks.  Subsequently, Fath
\cite{fath04_cyber} and Fath and Killian \cite{fath07_pyramids} found
evidence to support the network non-locality and homogenization
hypotheses in large-scale hypothetical model ecosystems built from an
ecosystem assembly algorithm; however, they found that network
amplification occurred only rarely.  Salas and Borrett
\cite{salas11_did} found that the network non-locality hypothesis held
in 74\% of 50 empirically-based trophic ecosystem networks, and
Borrett and Salas \cite{borrett10_hmg} showed that network
homogenization occurred in 100\% of the same 50 models.  Fath and
Patten \cite{fath98} provided an algebraic proof for why network
synergism should always occur in ecosystem models.  Thus, the existing
evidence supports the claim that pathway proliferation, network
non-locality, network homogenization, and synergism are general
properties of trophic-based ecosystem models, while network
amplification is an occasional property or possibly a modeling
artifact.  Network mutualism has rarely been investigated.

The majority of evidence for these ecosystem properties is based on
trophic ecosystem models.  We posit that each ecosystem model class
may have different characteristic network properties or intensities of
the properties.  That said, the ENA framework predicts that all six
described ecosystem properties are general across the model classes
because these models still represent an ecosystem at their core.

\subsection{Objectives}
We investigated whether these six ENA properties generally occur in
biogeochemical cycling ecosystem networks.  We specifically tested
two hypotheses. First, we hypothesized that the pathway proliferation,
non-locality, homogenization, amplification, mutualism, and synergism
properties tend to occur in the BGC models because the properties are
purportedly general to ecosystems. Second, we hypothesized that these six
properties would tend to have a greater intensity in BGC models than
in TRO models because (1) BGC models tend to have higher rates of
recycling, and (2) all of the properties tend to increase with
recycling \cite{patten90, fath04_cyber, salas11_did}.  If these
hypotheses hold, then this work extends the generality of the
hypothesized ecosystem properties across both the trophic and
biogeochemical cycling model classes, and provides evidence that
differences between model classes can be functionally significant.

\section{Materials and Methods}

\subsection{Ecosystem Network Selection}
\begin{figure}[t]
\center
\includegraphics[scale=1.25]{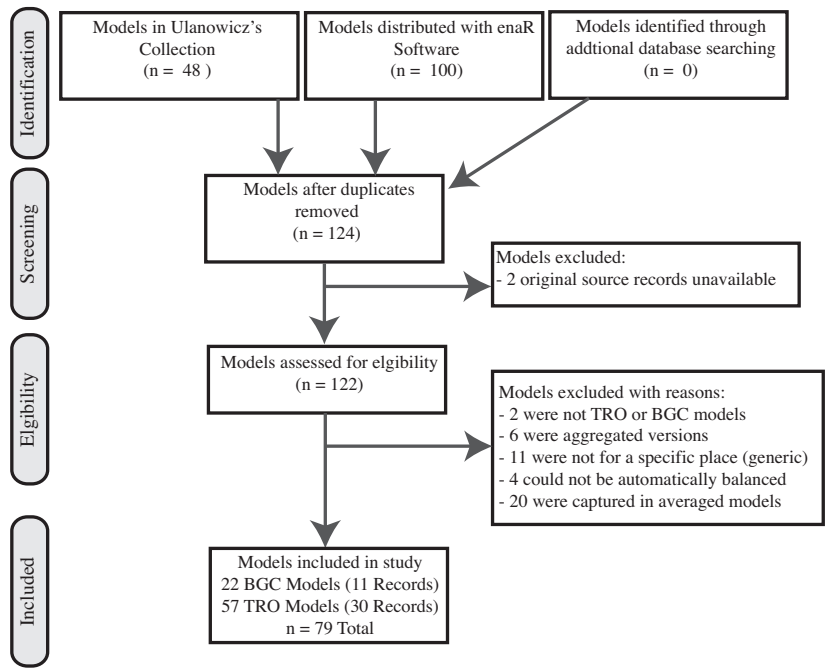}
   \caption{Information flow for ecosystem network model selection
     adapted from the PRISMA (Preferred Reporting Items for Systematic
     reviews and Meta-Analyses) guidelines.
     } \label{fig:prisma}
\end{figure}

Our approach to testing the generality of the ecosystem properties was
to apply ENA to a set of ecosystem network models and compare the
results.  As such, this study has characteristics of a systematic
review and meta-analysis, and we can characterize our model selection
for the study with an adaptation of the PRISMA (Preferred Reporting
Items for Systematic reviews and Meta-Analyses) guidelines
\cite{liberati09}. Figure~\ref{fig:prisma} provides an overview of
this decision process.

To identify ecosystem networks for this study, we started with two
existing collections.  The first set of 48 models was compiled
by Dr.\ Robert Ulanowicz
\url{http://www.cbl.umces.edu/~ulan/ntwk/network.html} over the course
of his career developing and promoting ENA.  The second is a
collection of 100 ecosystem models extracted from the literature and
distributed with the \texttt{enaR} R package for ENA
\cite{borrett14_enar}.  These were collected from colleagues and
periodic literature searches (2000 -- 2014).  Together, this identified
124 unique ecosystem networks.  In an initial screening step we
excluded two of these networks because the original model sources were
unavailable for review.

We then assessed the eligibility of the remaining 122 ecosystem
networks for our study.  To be included, the models must have met five
inclusion criteria. First, the models needed to fit the
characteristics of a trophic or biogeochemical cycling network.  We
excluded two models based on this criteria -- one was modeled a
network of water flow in a watershed, and another was for the Polish
economy.  Second, we included only the highest resolution (least
aggregated) networks for a particular system. This criteria is
necessary because some systems have been modeled multiple times with
differing network order (number of nodes used to partition the
system).  Third, to be included in our study, the model had to be
empirically-based in the sense that the investigators were attempting
to model a specific ecosystem and a portion of the model was
parameterized with empirical measurements from that system.  Fourth,
for analytical reasons (see next section) the models included had
to be at steady-state (inputs equal outputs).  Thus, we excluded models that
were not initially at steady-state that could not be automatically
balanced using \texttt{enaR}.  Fifth, there were several cases where
there were multiple models of the same system for different times or
conditions.  If an averaged model was also initially identified for
the site, we included only the averaged model to limit the bias of any
particular system. In some cases a similar conceptual model was
parameterized for different elements (e.g., C, N, P, Mg).  We included
this variety because we expected this to provide alternative views of
the same system.

Through this process we selected 79 ecosystem network models from 41
records.  This includes 22 biogeochemical cycling ecosystem networks
(Table~S1) and 57 trophic networks (Table~S2).  Eight distinct systems
are modeled in the set of biogeochemical network class and there are
40 separate systems in the trophic network class.  This set of trophic
models overlaps (49 of 50) with those analyzed for network
non-locality and network homogenization \cite{borrett10_hmg,
  salas11_did}.

\subsection{Ecological Network Analysis} \label{sec:ena}
While the specifics of the ENA algorithms have been presented in the
literature \cite{ulanowicz86, fath99_review, latham2006, schramski11}
and made available in software packages \cite{fath06,
  kazanci07, borrett14_enar}, the techniques are not well
known.  Further, in some cases we modified the traditional ENA network
statistics to facilitate their consideration across networks with
different numbers of nodes. Thus, this section describes the analyses
used in detail.

\subsubsection{Ecosystem Network Models}
Like network models of any complex system, ecosystem networks
represent the system with a set of nodes and edges
(Fig.~\ref{fig:example}).  Nodes represent biological species, groups
of species, or non-living components.  For example, a living node
could represent detritivores, which is a functional grouping of
species based on their eating habits, and non-living nodes might be
particulate organic matter (POM) or inorganic nitrogen in the form of
ammonium (NH\textsubscript{4}\textsuperscript{+}).  Edges are directed
and weighted. They indicate the flow of energy or matter from one node
$j$ to another $i$ over some time period, $\mathbf{F}=[f_{ij}]$
$i,j = 1, 2, ..., n$.  These networks allow only one edge from any
node to any other node, and self-loops may occur (e.g., cannibalism in
a food web).  These models typically focus on tracing a single
thermodynamically conservative tracer. Thus, an edge may capture the
effect of many physical, chemical, and biological processes including
feeding, excretion, death, and chemical conversions.

Ecosystems are thermodynamically open systems. This implies that the
systems have inputs and outputs from their environment, which are
modeled as a second class of directed edges.  Input edges, $z=[z_i]$
do not have a specified starting node and output edges $y=[y_j]$ have
no terminal node.  Alternatively we could assign a node to represent
the environment, but this is uncommon in the ENA literature.

Given an ecosystem network, we can apply ENA.  There
are several distinct analyses within ENA \cite{fath99_review,
  ulanowicz04}; we applied structural, flow, and utility analyses.

\subsubsection{Structural Analysis}
Structural analysis in ENA focuses on properties of the network
topology.  This topology is specified by the existence of connections
among the nodes; node and edge weights are ignored.  The analysis
also typically focuses on the within system connections, so the
boundary inputs and outputs are also ignored.  Thus, the analysis
primarily operates on the binary adjacency matrix, $\mathbf{A}=[a_{ij}]$,
induced from $\mathbf{F}$.  If there is a flow
from $j$ to $i$, $a_{ij} = 1$; otherwise, $a_{ij} = 0$.

A number of network statistics are used to characterize the digraph
structure.  The most common are the total number of nodes $n$ (graph
order), the number of edges or links $L = \sum\sum a_{ij}$, and
connectance $C=L/n^2$ (network density).  These statistics are a
necessary starting point, but they do not describe the pattern of
connections among the nodes.

Structural analysis is also concerned with features of the network
connectivity, which are characterized with graph theory and matrix
algebra \cite{berman79, brandes05}.  A strongly connected network is
one in which every node is reachable by a pathway of some length from
every other node.  Some networks can be decomposed into one or more
strongly connected components (SCC).  These SCCs are the largest
subset of nodes that are still reachable from each other
\cite{berman79}.  Most food webs and ecosystem networks are composed
of one or sometimes two SCCs with more than one node, which we call
large SCCs \cite{borrett07_jtb, borrett10_idd}. Often these networks
also have a set of additional nodes that are weakly connected to the
others; they are only reachable if we ignore the edge directions.  For
example, a plant node in a trophic network tracing carbon typically
has an edge showing consumption by one or more herbivores; however,
there is typically no input of carbon to the plants from within the
system since they take it up from the atmosphere.  We can quantify
these aspects of network connectedness by counting the number of large
($n>1$) SCCs in the network and the percent of network nodes that
participate in the SCCs.

Pathway proliferation appears when we count the number of pathways or
walks between any two nodes as we increment the pathway length $m$
(Fig.~\ref{fig:example}c) \cite{patten82_texoma, patten85}.
Raising the adjacency
matrix to the power $m$ returns a matrix $\mathbf{A}^m$ whose elements
$a^{(m)}_{ij}$ provide the walk count from $j$ to $i$.  In strongly
connected networks, $a^{(m)} \rightarrow \infty$ as $m \rightarrow
\infty$.

The rate of pathway proliferation varies among ecological networks
\cite{borrett03,borrett07_jtb}. This variation is due to the size of
the network, its direct connectance, and the topological arrangement
of the direct edges \cite{borrett03}.  The asymptotic rate of pathway
proliferation in an SCC is equivalent to the dominant eigenvalue of
the adjacency matrix associated with the SCC, $\ppr$
\cite{borrett07_jtb}.  If a network contains only one SCC, then the
pathway proliferation rate for the network and the SCC are identical.
$\ppr$ may also serve as a measure of the topological cyclicity in the
network \cite{fath07cyclic, fath07structural}.  As $\ppr$ scales
between 0 and $n$, we used a re-scaled pathway proliferation rate
$PPR=(\ppr )/n$ to compare the pathway proliferation rates among
networks of different sizes.  Notice that $0 \le PPR \le 1$.



\subsubsection{Flow Analysis}
Flow analysis characterizes the geometry of the ecological networks,
taking into consideration the flow weights \cite{hannon73,
  ulanowicz91}.  Further, it requires the boundary flux information.
Given its origination in economic Input-Output Analysis, flow analysis
can have either an input or output orientation; we focus on the output
orientation for this paper.  The input analysis typically generates
qualitatively similar results \cite{borrett11_ree}.

The ENA flow analyses we applied have four key steps.  The first step
is to calculate the node throughflows, which are a sum of the total
amount of energy or matter flowing into or out of each node.  This
can be calculated as

 \begin{align}
  T^{\textrm{input}}_i &= \sum_j^n f_{ij} + z_i \: (i,j = 1, 2,
                         \ldots, n)\textrm{, and} \\
  T^{\textrm{output}}_j &= \sum_i^n f_{ij} + y_j \: (i,j = 1, 2, \ldots, n).
 \end{align}
The flow analysis applied here assumes the network model is at
steady-state, which implies that $T_i^{\textrm{input}} =
T_j^{\textrm{output}}$ = $T_j$.

The second step in flow analysis is to calculate the direct flow
intensities among nodes ($\mathbf{G} = [g_{ij}]$). For the output
orientation, these values are defined as
\begin{align}
g_{ij} = \frac{f_{ij}}{T_j}. \label{eq:G}
\end{align}
Elements of $\mathbf{G}$ indicate the flow intensity from $j$ to $i$
over pathways of length $m = 1$.

The third step in our analysis is to find the integral flow intensity
matrix $\mathbf{N}=[n_{ij}]$. This matrix satisfies the following
equation.
\begin{align}
T = \mathbf{N} z \label{eq:T}
\end{align}
The $n_{ij}$ elements are effectively the sum of the flow intensities
from $j$ to $i$ across pathways of all lengths ($m = 0, 1, 2, 3,
\ldots, \infty$).  Thus, we can re-represent $\mathbf{N}$ as the sum
of the following series:
  \begin{align}
    \mathbf{N} &= \sum_{m=0}^\infty G^m \\ \label{eq:N}
    &= \underbrace{\mathbf{G}^0}_{\textrm{Boundary}} + \underbrace{\mathbf{G}^1}_{\textrm{Direct}} + \underbrace{\mathbf{G}^2 + \ldots + \mathbf{G}^m
      + \ldots}_{\textrm{Indirect}}.
  \end{align}
The elements of $\mathbf{G}^m$ indicate the flow intensity from $j$ to
$i$ over all pathways of length $m$.  Notice that when equation
 \ref{eq:N} is substituted back into equation \ref{eq:T}, the elements
 of $\mathbf{G}^0 z$ map the boundary inputs into $T$, $\mathbf{G}^1 z$
are the direct flows, and $\mathbf{G}^m z$ where $m>1$ are the indirect flows.
Given that ecosystems are thermodynamically open and the specification
of our model, this infinite series is convergent.  Therefore, we can
find $N$ using the identity $(\mathbf{I} - \mathbf{G})^{-1}$, where
$\mathbf{I}$ is the identity matrix.

From this primary flow analysis, we derive several whole-network
statistics relevant to this paper.  The first two are based on the
throughflow vector.  Total system throughflow (TST) is the sum of the
node throughflows ($TST = \sum T_j$) and is an indicator of the size
or activity of the system \cite{patten76, finn76}.  Average path
length ($APL = TST/\sum z_i$) indicates the average amount of activity
generated by a unit of input into the system \cite{finn76}.  This is
similar in concept to the multiplier effect in economics
\cite{samuelson48} and has been termed network aggradation in ecology
\cite{jorgensen00}.  The third indicator we used is the Finn Cycling
Index (FCI), which indicates the fraction of throughflow derived
from recycling as opposed to acyclic flows \cite{finn76, finn80}.

The remaining flow-based network statistics quantify three of the
remaining ENA properties of interest (Table~\ref{tab:stats}).  Network
non-locality is characterized by the ratio of indirect-to-direct flows
in the network ($\ID$); when $\ID > 1$, indirect effects are said to
dominate direct effects and network non-locality is present.  Network
homogenization is a ratio of the coefficient of variation in the
direct flow intensity matrix to the coefficient of variation in the
integral flow matrix ($HMG$).  When $HMG > 1$ the flows are more
evenly distributed (lower variance) in the integral than in the flow
matrix, which implies that the indirect flows are more evenly
distributing the system activity across the network and network
homogenization is present.  Network amplification occurs when an
off-diagonal element in the integral flow matrix exceeds 1, which
indicates that the flow is amplified from the initial boundary input.
The strength of amplification ($AMP$) is quantified by counting the
number of non-diagonal elements are greater than one. To allow
comparison across network models of different sizes, we divide this
count by the total number of possible off diagonal elements.

\begin{table}[t]
  \caption{Formulation of six network statistics to characterize
    hypothesized general properties of ecosystems.} \label{tab:stats}
  \center
  \begin{tabular}{l l l l}
    \hline ENA Property & Symbol & Statistic \\ \hline \\
    Pathway Proliferation & PPR & $\lambda_1(\mathbf{A})/n$ \\ [2 ex]
    Network Non-locality & \ID &
    $\sum\left(\left(\mathbf{N}-\mathbf{I}-\mathbf{G}\right)
      z\right) / \sum\mathbf{G}z$\\ [2 ex]
    Network Homogenization & $HMG$ & $CV(\mathbf{G})/CV(\mathbf{N})$\\ [2 ex]
    Network Amplification  & $AMP$ & $ \left(\# n_{ij \: i\neq j} >1
    \right)/\left( n(n-1) \right)$\\ [2 ex]
    Network Mutualism & $MUT$ & $\left( \# \upsilon_{ij} > 0 \right) /
    \left(\# \upsilon_{ij} < 0 \right)$ \\ [2 ex]
    Network Synergism & $SYN$ & $\left(\sum\sum \upsilon_{ij} \in
      \mathbb{R}_{>0}\right) / \left(|\sum\sum
      \upsilon_{ij} \in \mathbb{R}_{<0} | \right)  $ \\
\hline
\end{tabular}
\end{table}

\subsubsection{Utility Analysis}
The network mutualism and network synergism properties appear
from the application of ENA utility analysis.  This technique
determines the net relationships among nodes in the network that are
derived from the direct transactions \cite{patten91,fath98,
  fath07_mut}.  Here, we use the flow based utility analysis
\cite{fath06, borrett14_enar}.  The first step in this analysis is to
calculate the direct utility matrix as
\begin{align}
  \mathbf{D} = [d_{ij}] = \frac{f_{ij}-f_{ji}}{T_i}, \quad i,j = 1, 2,
  \ldots, n \label{eq:D}
\end{align}
such that $-1 \le d_{ij} \le 1$.  Further, $d_{ij}$ shows the direct
utility intensity from $i$ to $j$ over pathways of length $m = 1$.  As
in flow analysis, the integral utility intensity matrix is then computed as
  \begin{align}
    \mathbf{U} &= \sum_{m=0}^\infty D^m\\ \label{eq:U} &=
    \underbrace{\mathbf{D}^0}_{\textrm{Boundary}} +
    \underbrace{\mathbf{D}^1}_{\textrm{Direct}} +
    \underbrace{\mathbf{D}^2 + \ldots + \mathbf{D}^m +
      \ldots}_{\textrm{Indirect}}.
  \end{align}
The elements of $\mathbf{U}$ indicate the integral utility intensity
from $i$ to $j$.  When the dominant eigenvalue of $\mathbf{D}$ is less
than one, this power series converges such that $\mathbf{U} =
(\mathbf{I} - \mathbf{D})^{-1}$.  When the convergence criteria fails,
the identity relationship should not be mathematically appropriate.
Despite this mathematical problem, use of the identity has been common
in the literature and it appears to return ecologically meaningful
results. Why this appears to work despite the mathematical issue is an
open problem in ENA.  As such, in this paper we complete the utility
analysis even when the convergence criteria fails, but identify the
results as mathematically suspect.

The next step in utility analysis is to re-scale and dimensionalize
both the direct and integral utility matrix by multiplying them by the
node throughflows,
\begin{align}
\mathbf{\Delta} &= [\delta_{ij}] = \breve{\mathbf{T}}
\mathbf{D}, \quad \textrm{and}  \label{eq:delta} \\
\mathbf{\Upsilon} &= [\upsilon_{ij}] = \breve{\mathbf{T}}
\mathbf{U},  \label{eq:upsilon}
\end{align}
where $\breve{\mathbf{T}}$ is a $n \times n$ matrix with the node
throughflows on the principle diagonal and zeros elsewhere.  Thus,
$\delta_{ij}$ is the dimensionalized [M L\textsuperscript{-2}
T\textsuperscript{-1}] direct utility $i$ receives from $j$, and
$\upsilon_{ij}$ is the dimensionalized integral utility.

The tests for network mutualism and network synergism involve the
ratios of positive and negative utilities in $U$, which have been
termed the benefit--cost ratios (Table~\ref{tab:stats}).  In both
cases, the property is said to occur if the ratios are greater than 1,
indicating that the positive utilities exceed the negative utilities.

\subsection{Data Analysis}
Our primary data analysis for this paper had three main components.
First we used the \texttt{enaR} package to apply ENA to the selected
BGC models to test the hypothesized generality of the properties in
the biogeochemical cycling ecosystem networks.  We evaluated the
presence of each ecosystem property as expressed in the network
statistics described (Table~\ref{tab:stats}).  Second, we used a
two-sample Wilcoxan signed rank test to compare the intensity of each
ecosystem property between the BGC and TRO sets of models
(significance criterion $\alpha = 0.05$).  Third, to test the effect
of model size and recycling on the ENA properties, we applied the
non-parametric Spearman Rank correlation.

\subsection{Uncertainty Analysis}
To evaluate the effect of flux uncertainty on the ENA results, we
applied an inverse linear modeling technique \cite{vezina1988,
  kones09}.  We used the limSolve \R\ package \cite{soetaert09} to
generate an ensemble of 10,000 plausible models for each network by
simultaneously but independently perturbing the model fluxes by a
random amount up to $\pm 50\%$ drawn from a uniform distribution.
These perturbed models were constrained to meet the steady-state
assumption of ENA.  We then applied ENA to the ensemble of models and
calculated the 95\% confidence interval for each network statistic.

\section{Results}
The results of this study generally support the hypotheses.  We
report the results in four stages.  We first present the results
of the ENA network statistics in the biogeochemical cycling models,
and then we show the values for the trophic based models.  The third
subsection compares the distribution of these properties between the
two model classes.  In part four, we confirm the importance of cycling
intensity for these six ecosystem properties.

\subsection{ENA properties in Biogeochemical Cycling  Ecosystem Networks}
\begin{figure}[htp!]
\center
\includegraphics[scale=0.78]{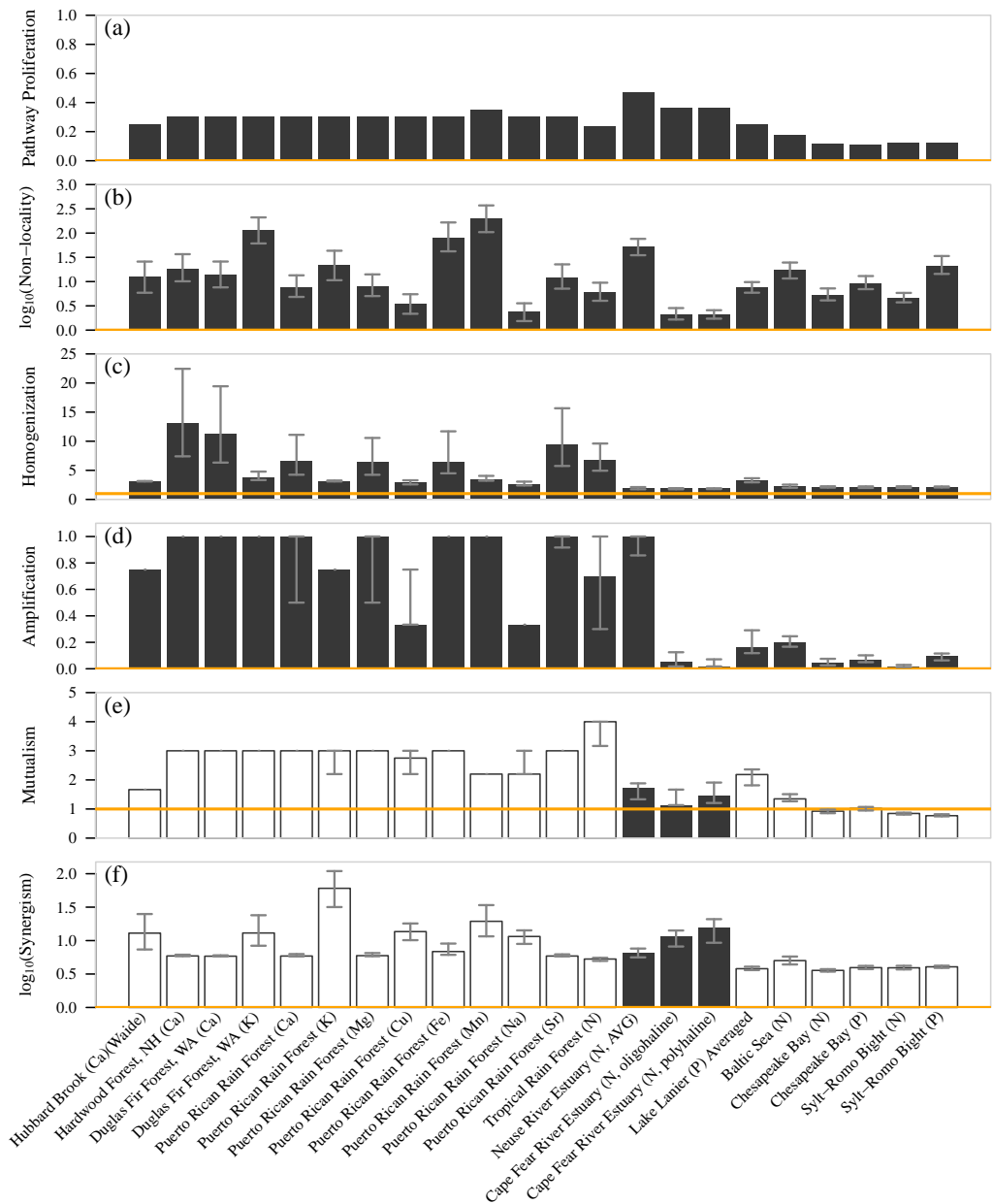}
  \caption{Intensity of the (a) relative pathway proliferation rate
    ($PPR$) (b) network non-locality ($I/D$), (c) network
    homogenization ($HMG$), (d) network amplification ($AMP$), (e)
    network mutualism ($MUT$), and (f) network synergism ($SYN$) in 22
    biogeochemical cycling ecosystem networks.  Notice that
    non-locality and synergism were $\log_{10}$ transformed to
    facilitate comparisons (recall that $\log_{10}(1) = 0$).  Models
    are ordered from smallest (left) to largest (right) $n$, and the
    error bars represent the 95\% confidence interval based on the
    uncertainty analysis.  The open bars in network mutualism and
    network synergism indicate that the utility analysis failed to
    meet the convergence criteria.} \label{fig:ep-bgc}
\end{figure}

Pathway proliferation, non-locality, homogenization, amplification,
and synergism occurred to varying degrees in all 22 BGC models analyzed
(Fig.~\ref{fig:ep-bgc}). Network mutualisms occurred 86\% of the
models. Thus, the results support the first hypothesis that these are
general properties of ecosystems.

The relative rate of pathway proliferation ranged from 0.11 to 0.47
(Fig.~\ref{fig:ep-bgc}a).  This proliferation rate reflects the results
for the whole network because the models each contain a single large
($n>1$) strongly connected component. Furthermore, all of the nodes
participate in the model's large SCC except in the Hubbard Brook Ca
model in which only 3/4 do.  As this is purely a topological property,
the flux uncertainty analysis did not affect these results.

In all of the BGC, models $I/D > 1$ ($log_{10}(I/D) > 0$).  This
indicates the dominance of indirect flows and the presence of network
non-locality.  The median realized $I/D$ was 12.2 in the BGC models,
but it ranged between 2.07 in the polyhaline model of N cycling in the
Cape Fear River Estuary to 201.1 in the Mn cycling model in the Puerto
Rican rain forest model.  The 95\% CI intervals from the uncertainty
analysis (error bars) indicate a range of uncertainties for the
specific values for each network, but none of these intervals cross
the interpretation threshold of $I/D = 1$.

Likewise, the homogenization statistic exceeded one in all BGC models
($HMG > 1$).  This indicates that the integral flow intensity matrices
were less variable than the direct flow intensity matrices.  When the
pathways of increasing length were considered, the particular nutrient
being traced through the network was more evenly distributed between
the nodes. $HMG$ ranged from 1.84 to 13.  Again, the 95\% CI do not
cross the interpretation threshold of $HMG = 1$.

Network amplification occurred ($AMP>0$) in all 23 BGC models; the
median value faction of possible off-diagonal positions with elements
greater than 1 was 0.75.  The lowest level of amplification was 2\% in
the polyhaline Cape Fear Estuary model of the N cycle and
amplification peaked at 100\% in nine of the networks.  The 95\% CIs
reinforced this result.

The network mutualism statistic ranged from 0.77 to 4 in the BGC
models, where 86\% of them had a value larger than one.  Only the
Chesapeake Bay N model and the Sylt-R{\o}m{\o} Bight N and P models
failed to exhibit the network mutualism property.  Notice, however,
that the 95\% uncertainty interval for the Chesapeake Bay N model
includes one.  As expected all of the models exhibited network
synergism. The synergism statistic ranged from 0.55 to 1.78 on a log
scale.  However, the Utility matrix failed the convergence criterion
($\lambda_1(\mathbf{D}) < 1$) in all but three models, making these
results mathematically suspect (Fig.~\ref{fig:ep-bgc}).

\subsection{ENA Properties in Trophic Ecosystem Networks}
\begin{figure}[htp!]
\center
\includegraphics[scale=0.8]{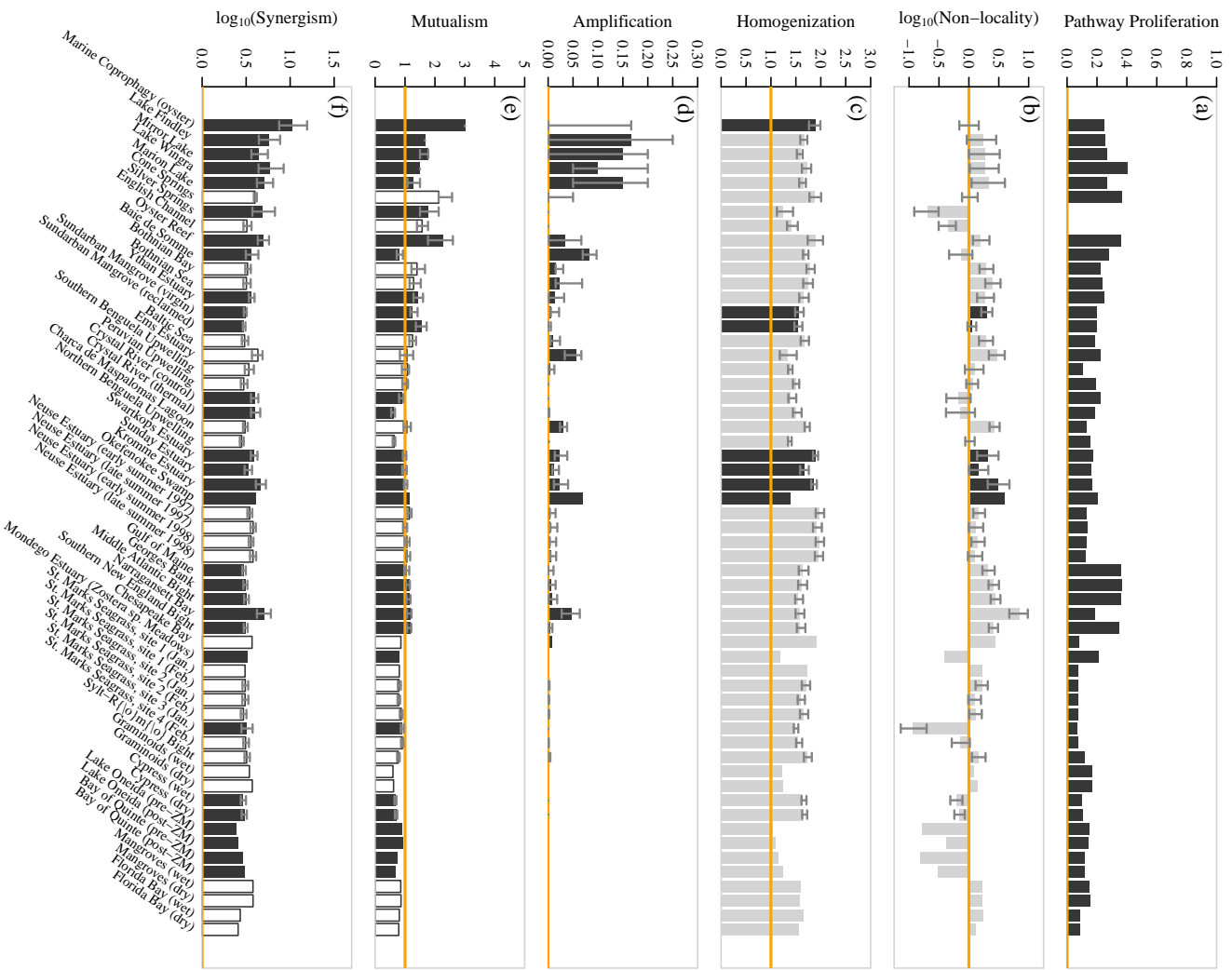}
  \caption{Intensity of the (a) relative pathway proliferation rate
    ($PPR$), (b) network non-locality $I/D$), (c) network
    homogenization ($HMG$), and (d) network amplification ($AMP$), (e)
    network mutualism ($MUT$), and (f) network synergism ($SYN$) in 57
    trophic ecosystem networks.  Non-locality and network synergism
    were $\log_{10}$ transformed to facilitate comparisons.  Models
    are ordered from smallest (left) to largest (right) $n$.  Light
    gray bars indicate results previously reported
    \cite{borrett10_hmg, salas11_did}, and the open bars in network
    mutualism and network synergism indicate that the utility analysis
    failed to meet the convergence criteria for these models.} \label{fig:ep-tro}
\end{figure}

There is also general support for five of the six ENA hypotheses in
the 57 TRO models (Fig.~\ref{fig:ep-tro}), though these results are
tempered by the uncertainty analysis.  Note that the uncertainty
analysis technique failed to build an ensemble of perturbed models for 14 of
the TRO models due to ``incompatible constraints'' that made the
solutions impossible to identify \cite{vezina1988}, including the eight
largest models.

Pathway proliferation occurred in all but two models: the Silver
Springs and the English Channel.  The maximum relative pathway
proliferation rate was 0.4.  Non-locality occurred in 75.4\% of the
initial models, including all 4 of the models not previously
considered. However, with the 50\% flux uncertainty the 95\% CI of the
non-locality statistics for 14 of the models crossed the existence
threshold, which leaves us uncertain of their status.  Network
homogenization occurred in all 57 TRO models with values ranging from
1.04 to 1.96.    Network amplification was found in 75\% of the trophic
models, with values ranging from 0.01\% to 17\%.  Network synergism
occurred in all 57 models with $log_{10}$ transformed parameter values
ranging from 0.38 to 1.02.  The 95\% CI for network homogenization,
amplification, and synergism all exceeded the exceeded the property
existence thresholds.

The evidence for the generality of network mutualism was not strong;
the network statistic ranged from 0.60 to 3 in the TRO models.  The
network mutualism parameter only exceeded one, indicating the ENA
property presence, in 28 of the 57 models (49\%).  Further, the
uncertainty analysis showed that the evidence was inconclusive for 7
of these 28 models. Also notice that 47.4\% of the TRO models failed to
meet the convergence criterion for the utility power series, which
makes these results mathematically suspect.

The results for non-locality and homogenization are not surprising
given the known results on 49 of these models
\cite{borrett10_hmg, salas11_did}.  However, these two properties
were found in all seven of the unique models in this set.

\subsection{BGC versus TRO network models}
\begin{figure}[t!]
\center
\includegraphics[scale=0.8]{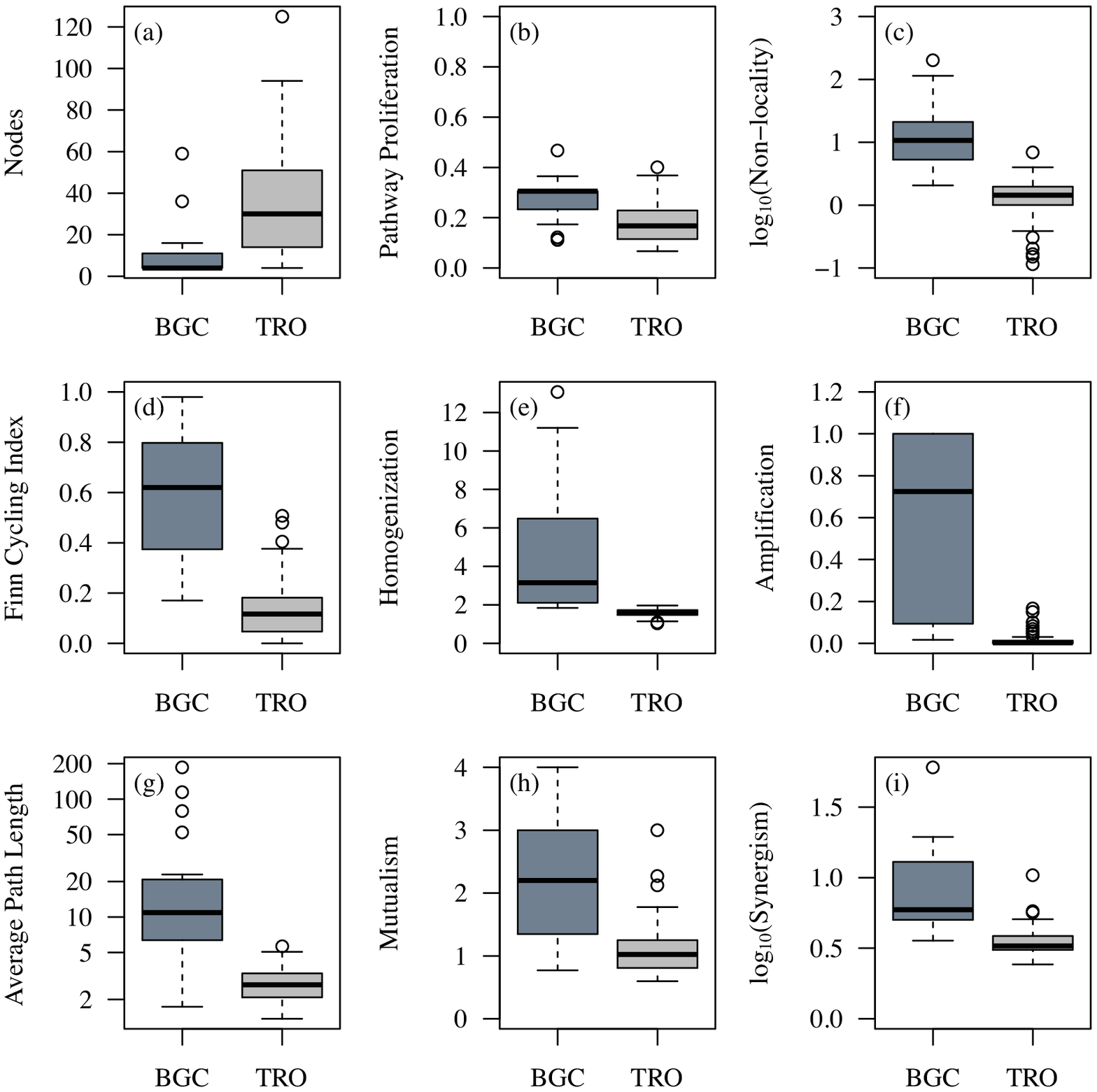}
  \caption{Comparison of nine network properties in a set of
    biogeochemical cycling (BGC, N = 22) and trophic (TRO, N = 57)
    ecosystem networks.  These properties include (a) number of nodes,
    (b) pathway proliferation rate, (c) network non-locality, (d) Finn
    Cycling Index, (e) network homogenization, (f) network
    amplification, (g) average path length, (h) network mutualism, and
    (i) network synergism. } \label{fig:comp}
\end{figure}

When we compared the distributions of the ENA statistics between the
two classes of ecosystem models (BGC vs TRO), we found a significant
difference for each metric (Fig.~\ref{fig:comp}).  As expected from
our class definitions, these distributions indicate that the BGC
models in our data set tended to have fewer nodes, connectance, and a
higher degree of recycling.  They also tended to have a higher average
path length.  Further, as we hypothesized the relative pathway
proliferation rate, network non-locality, network homogenization,
network amplification, network mutualism, and network synergism tended
to be greater in the BGC models.  A non-parametric Wilcoxan rank sum
test showed a statistically significant difference between BGC and TRO
models for the ten metrics compared (Table~\ref{tab:comp-stat}).  The
p-values are all below the critical value of 0.05, which indicates
that there is sufficient evidence to reject the statistical null
hypotheses of no difference.

\begin{table}[h]
\caption{Summary of Wilcox Rank Sum statistical comparison of the
  distribution of ten network statistics between BGC and TRO
  models.} \label{tab:comp-stat}
\center
\begin{tabular}{lrrl}
  \hline
Network Statistic & W & p-value & Significance\\
  \hline
  $n$ & 240 & 1.28e-05 & *\\
  $C$ & 912 & 3.92e-03 & *\\
  $FCI$ & 1,218 & 6.05e-10 & *\\
  $APL$ & 1,118 & 3.22e-07 & *\\
  $PPR$ & 985 & 2.35e-04 & *\\
  $\ID$ & 1,243 & 1.06e-10 & *\\
  $HMG$ & 1,260 & 3.09e-11 & *\\
  $AMP$ & 1,234 & 1.72e-10 & *\\
  $MUT$ & 1,041 & 5.86e-06 & * \\
  $SYN$ & 1,163 & 4.72e-06 & * \\
   \hline
\end{tabular}
\end{table}

\subsection{Importance of Cycling}
\begin{figure}[ht!]
\center
\includegraphics[scale=1]{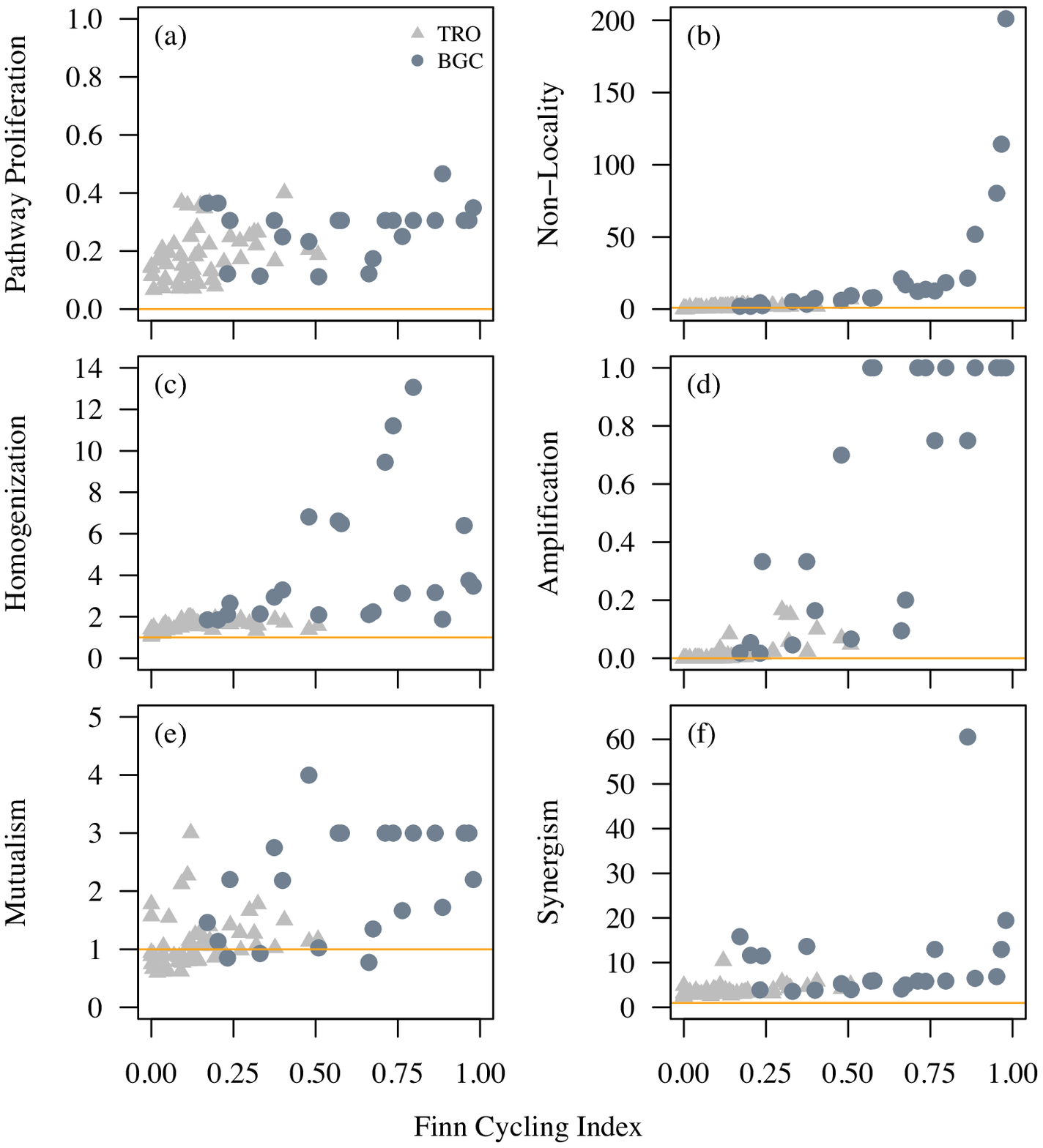}
\caption{Scatter plots showing the relationship between (a) relative
  pathway proliferation rate, (b) network non-locality ($I/D$), (c)
  network homogenization (HMG), and (d) network amplification (AMP) in
  79 ecosystem network models.} \label{fig:cyc}
\end{figure}

We hypothesized that the explanation for why the ENA properties would
be more intense in the BGC models was due to their tendency to have
greater cycling.  Figure~\ref{fig:cyc} confirms for this data set that
all six properties tend to increase with the Finn Cycling Index,
though the shape and intensity of the relationship varies.  The
correlations between FCI and the ENA property indicators was
statistically significant (Table~\ref{tab:cor}).

\begin{table}[ht]
  \center
  \caption{Results of a Spearman rank correlation test between the
    Finn Cycling Indices and six Ecosystem Network Analysis statistics
    for 79 ecosystem models.} \label{tab:cor}
\begin{tabular}{lrrrl}
  \hline
Network Statistic & rho & t & p-value & Significance \\
  \hline
PPR & 0.51 & 37588.97 & 2.69e-06 & * \\
  I/D & 0.94 & 5211.53 & 8.40e-37 & * \\
  HMG & 0.74 & 21543.63 & 8.77e-15 & * \\
  AMP & 0.93 & 5811.33 & 4.85e-35 & * \\
  MUT & 0.62 & 31595.33 & 1.58e-09 & * \\
  SYN & 0.71 & 23824.64 & 2.36e-13 & * \\
   \hline
\end{tabular}
\end{table}

\section{Discussion}

\subsection{Ecological Significance}
Ecosystem network analysis and its associated environ theory have been
developing since the mid 1970s \cite{hannon73, patten76,
  fath99_review, borrett12_netecol}.  Today, these techniques and
theory appear to be in a maturation phase.  This work makes two
primary contributions to this maturing systems ecology understanding
of ecosystems.

First, the results presented here extend the evidence in support of
five of the hypothesized general ecosystem properties: pathway
proliferation, network non-locality, network homogenization, network
amplification, and network synergism.  Theses properties tend to occur
in empirically-based network models of two classes:  trophic and
biogeochemical cycling.  Network mutualism was more variable,
especially in the BGC models, and strength of network amplification
did decline with network size as it did in simulated ecosystem networks
\cite{fath04_cyber}.  This evidence adds to the work from
studies of empirically-based ecosystem networks \cite{patten90_cycles,
  salas11_did, fath2013dependence} as well as generative ecosystem
models \cite{fath04_cyber, fath07_pyramids}. Further, these results
were generally (but not always) robust to the uncertainty introduced
into the flux estimates.

Second, the results suggest that the lens used to view the ecosystem
-- trophic or biogeochemistry -- can influence the intensity of the
observed network properties. In general, models with more cycling tend
to have a more intense expression of these properties.  This concern
with analytical point of view is similar to Reiners' \cite{reiners86}
recognition of multiple complimentary models for ecosystems.

When considered in aggregate, these six hypothesized general
properties of ecosystems reveal some of the ecological consequences of
ecosystem connectivity through the lens of a single transactive
currency. The intensity and pattern of direct connections among
ecosystem members leads to indirect interactions that alter the
ultimate nature of the interactions.

\subsection{Limitations and Opportunities}

A limitation of this study concerns the two sets of ecosystem
network models.  While the models themselves likely vary in quality,
there are fewer BGC models in our sample than TRO models and the BGC
models tended to have fewer nodes. This likely effects our inference
about the differences between the two network classes; however, we
expect the general trends reported in this paper to hold.  This
claim is supported by the general increase in the ecosystem network
properties in the nitrogen and phosphorus models when compared to the
related carbon models of both the Chesapeake Bay and Sylt-R{\o}m{\o}
Bight models (Figs~\ref{fig:ep-bgc}-\ref{fig:ep-tro}). 

To determine the robustness of our results to potential
parameter uncertainty, we applied an uncertainty analysis based on
inverse linear modeling \cite{vezina1988, kones09, soetaert09}.  While
this kind of global uncertainty analysis \cite{saltelli2008global} is
useful for estimating our confidence in the results, it has some
limitations.  First, the technique assumes the model parameters are
independent when we might expect that some of the parameters may have
natural inter-dependencies.  Because the technique misses the
potential covariances, the technique as applied is conservative,
leading to potential type II statistical inference errors. Second, in
the absence of better parameter uncertainties, we applied a uniform
$\pm$50\% parameter perturbation.  In reality, some of the parameters
may be more and less certain \cite{hines15}.  The effect of this
heterogeneity of natural variability and uncertainty is an issue that
needs further exploration.

While this second issue could increase or decrease the estimated 95\%
confidence intervals, we expect that when combined with the first
issue that the confidence intervals are over estimates of the true
uncertainty.  If true, this bolsters the evidence presented in this
paper for the support of the six ecosystem properties. A third issue
that will affect the potential broad applicability of the inverse
linear modeling based uncertainty analysis is the occasional failure
of the procedure to identify models that meet the mathematical
constraints.  This is a known challenge for the technique
\cite{vezina1988}.

Despite the methodological challenges, uncertainty analyses are
essential for making informed inferences from ENA
\cite{borrett07_lanier, kones09, kaufman10}.  Thus, we expect that
uncertainty analyses of some form will become standard practice for
future ENA work.

\subsection{Network Science}

Network models and analyses are useful in a wide range of disciplines
including medicine and genetics \cite{ennett1993peer,zhao2011},
ecology \cite{dunne02,anderson2011, borrett14_rise}, sociology
\cite{wasserman94}, education \cite{martinez2003}, and business
\cite{davis1996, wuellner2010, huang2011}. From these domain specific
theoretical and methodological developments, a network science is
emerging \cite{newman2010, brandes13, nrc06network}.  Brandes et al.\
\cite{brandes13} define network science broadly as the ``study of the
collection, management, analysis, interpretation, and presentation of
relational data'' as used to study a wide range of complex systems.
While some expect that this approach will identify properties,
structuring forces, and organizational consequences of connectance
common to complex systems in general \cite{barabasi12}, Brandes et
al.\ \cite{brandes13} claim a more modest goal for network theory.
They suggest that network theory will focus on how to abstract systems
into useful network representations and suggest appropriate ways of
querying these models.

Ecosystems have all of the hallmarks of complex adaptive systems
\cite{levin98, patten02_cahs}. This paper highlights a network model
to investigate ecosystem energy and matter transactions and what
appear to be common properties of this system class.  We suspect that
this has a broader relevance for network science.  This abstraction
and subsequent analysis was initially built upon economic Input-Output
Analysis \cite{hannon73, leontief65}, but has developed substantially
to meet the demands of ecological and environmental questions.  Thus,
this ecosystem network model and analysis maybe relevant for other
kinds of complex systems that can be modeled similarly.

The network science type proposition that ecosystem network models can
be divided into two (or more) distinct classes is pivotal for the work
presented in this manuscript.  The evidence in Fig.~\ref{fig:comp}
suggests that the models generally follow the expected differences
between the proposed classes.  Further, the hypothesized systems
properties do seem to be general in both groups.  Other possible
classes include hydrologic models \cite{patten82_ok} and urban
metabolism or industrial material flow models
\cite{bailey2004applyingI, zhang10, li12, chen12}.

Unfortunately, the distinction between TRO and BGC models is not
always crisp.  For example, the core conceptual model for the
Sylt-R{\o}m{\o} Bight TRO ecosystem was only slightly modified to
build the BGC models. For example, the node representing dissolved
organic carbon (DOC) was re-purposed to represent dissolved nitrogen,
and then the edge weights were estimated for the nitrogen fluxes
rather than the carbon fluxes.  Thus, the model topology is still
largely that of the original trophic model, while the geometry
responds to the different currency. In many cases this kind of
transition makes sense because multiple nutrients (currencies) travel
together in some ecosystem transactions.  For example, when deer eat
grass they obtain energy, carbon, and nutrients like nitrogen and
phosphorus through the same process, though the stoichiometric ratios
may vary in time and space \cite{sterner2002,
  lauridsen2014consequences}.  Thus, some of the TRO and BGC models
are not strictly independent, and perhaps in the future they could be
combined through the multilayer network framework recently introduced
\cite{kivela2014}.  Despite this issue, we assert that the ecosystem
model classification is a useful analytical technique that can capture
important differences among the ecosystem model types.  We look
forward to the extension of these ideas to other classes of ecosystem
models like the urban metabolism models \cite{kennedy2011study,
  chen12, li12} or other kinds of input-output systems
\cite{layton2012correlation, wiedmann2013}.

\section*{Funding}
This work was supported by the University of North Carolina Wilmington
and the National Science Foundation [DEB-1020944].

\section*{Acknowledgments}
This research and manuscript benefited from critiques by colleagues in
the Systems Ecology and Ecoinformatics Laboratory at UNCW.

\section*{Supporting Information}

\subsection*{S1 Table} \label{tab:TRO}
{\bf Twenty-two empirically-based ecosystem network models of
  biogeochemical cycling}

\subsection*{S2 Table} \label{tab:BGC}
 {\bf Fifty-seven empirically-based trophic ecosystem network}


\section*{Supporting Information}

\subsection*{S1 Table} \label{tab:TRO}
{\bf Twenty-two empirically-based ecosystem network models of
  biogeochemical cycling}

\subsection*{S2 Table} \label{tab:BGC}
 {\bf Fifty-seven empirically-based trophic ecosystem network}

\section*{Acknowledgments}
This research and manuscript benefited from critiques by colleagues in
the Systems Ecology and Ecoinformatics Laboratory at UNCW.

\end{spacing}

\nolinenumbers


\newpage

\begin{table}[ht]
\renewcommand\thetable{S1}
\caption{Twenty-two empirically-based ecosystem network models of
  biogeochemcial cycling.} \label{tab:BGC}
\begin{center}
\begin{tiny}
\begin{tabular}{lrrrrrrr}
  \hline
Model & Units & $n$ & $C$ & $Input$ & $TST$ & $FCI$ & Reference \\
  \hline
Hubbard Brook (Ca)(Waide) & kg Ca Ha$^{-1}$  yr$^{-1}$ &   4 & 0.25 &  11 & 168 & 0.76 & \cite{waide74} \\
  Hardwood Forest, NH (Ca) & kg Ca Ha$^{-1}$  yr$^{-1}$ &   4 & 0.31 &  11 & 200 & 0.80 & \cite{jordan72} \\
  Duglas Fir Forest, WA (Ca) & kg Ca Ha$^{-1}$  yr$^{-1}$ &   4 & 0.31 &   4 &  54 & 0.74 & \cite{jordan72} \\
  Duglas Fir Forest, WA (K) & kg K Ha$^{-1}$  yr$^{-1}$ &   4 & 0.31 &   0 &  45 & 0.97 & \cite{jordan72} \\
  Puerto Rican Rain Forest (Ca) & kg Ca Ha$^{-1}$  yr$^{-1}$ &   4 & 0.31 &  43 & 274 & 0.57 & \cite{jordan72} \\
  Puerto Rican Rain Forest (K) & kg K Ha$^{-1}$  yr$^{-1}$ &   4 & 0.31 &  20 & 433 & 0.86 & \cite{jordan72} \\
  Puerto Rican Rain Forest (Mg) & kg Mg Ha$^{-1}$  yr$^{-1}$ &   4 & 0.31 &  10 &  70 & 0.58 & \cite{jordan72} \\
  Puerto Rican Rain Forest (Cu) & kg Cu Ha$^{-1}$  yr$^{-1}$ &   4 & 0.31 &   0 &   2 & 0.37 & \cite{jordan72} \\
  Puerto Rican Rain Forest (Fe) & kg Fe Ha$^{-1}$  yr$^{-1}$ &   4 & 0.31 &   0 &   7 & 0.95 & \cite{jordan72} \\
  Puerto Rican Rain Forest (Mn) & kg Mn Ha$^{-1}$  yr$^{-1}$ &   4 & 0.38 &   0 &   7 & 0.98 & \cite{jordan72} \\
  Puerto Rican Rain Forest (Na) & kg Na Ha$^{-1}$  yr$^{-1}$ &   4 & 0.31 &  64 & 140 & 0.24 & \cite{jordan72} \\
  Puerto Rican Rain Forest (Sr) & kg Sr Ha$^{-1}$  yr$^{-1}$ &   4 & 0.31 &   0 &   1 & 0.71 & \cite{jordan72} \\
  Tropical Rain Forest (N) & g N m$^{-2}$ d$^{-1}$  &   5 & 0.24 &  10 &  71 & 0.48 & \cite{edmisten70} \\
  Neuse River Estuary (N, AVG) & mmol N m$^{-2}$ season$^{-1}$ &   7 & 0.45 & 795 & 41,517 & 0.89 & \cite{christian03} \\
  Cape Fear River Estuary (N, oligohaline) & nmol N cm$^{-3}$ d$^{-1}$ &   8 & 0.36 & 3,802 & 7,088 & 0.20 & \cite{hines12} \\
  Cape Fear River Estuary (N, polyhaline) & nmol N cm$^{-3}$ d$^{-1}$ &   8 & 0.36 & 3,068 & 5,322 & 0.17 & \cite{hines15} \\
  Lake Lanier (P) Averaged & mg P m$^{-2}$ day$^{-1}$ &  11 & 0.21 &  95 & 749 & 0.40 & \cite{borrett07_lanier} \\
  Baltic Sea (N) & mg N m$^{-3}$ day$^{-1}$  &  16 & 0.15 & 2,348 & 44,510 & 0.67 & \cite{hinrichsen98_baltic} \\
  Chesapeake Bay (N) & mg N m$^{-2}$ yr$^{-1}$  &  36 & 0.12 & 73,430 & 484,325 & 0.33 & \cite{baird95} \\
  Chesapeake Bay (P) & mg P m$^{-2}$ yr$^{-1}$ &  36 & 0.12 & 9,402 & 101,092 & 0.51 & \cite{ulanowicz99} \\
  Sylt-R{\o}m{\o} Bight (N) & mg N m$^{-2}$ yr$^{-1}$  &  59 & 0.09 & 99,613 & 363,693 & 0.23 & \cite{baird08_sylt} \\
  Sylt-R{\o}m{\o} Bight (P) & mg P m$^{-2}$ yr$^{-1}$  &  59 & 0.09 & 2,510 & 57,708 & 0.66 & \cite{baird08_sylt} \\
   \hline
\end{tabular}
\end{tiny}
\end{center}
\end{table}





\clearpage
\newpage
\begin{table}[ht]
\renewcommand\thetable{S2}
\caption{57 empirically-based, trophic ecosystem
  networks.} \label{tab:TRO}
\begin{center}
\begin{tiny}
\begin{tabular}{lrrrrrrr}
  \hline
Models & Units & $n$ & $C$ & $Input$ & $TST$ & $FCI$ & Reference \\
  \hline
Marine Coprophagy (oyster) & kcal m$^{-2}$ yr$^{-1}$  &   4 & 0.25 & 379 & 549 & 0.12 & \cite{haven66} \\
  Lake Findley  & gC m$^{-2}$ yr$^{-1}$  &   4 & 0.38 &  21 &  50 & 0.30 &  \cite{richey78} \\
  Mirror Lake & gC m$^{-2}$ yr$^{-1}$  &   5 & 0.36 &  72 & 217 & 0.32 &   \cite{richey78} \\
  Lake Wingra & gC m$^{-2}$ yr$^{-1}$  &   5 & 0.40 & 478 & 1,517 & 0.40 &  \cite{richey78} \\
  Marion Lake & gC m$^{-2}$ yr$^{-1}$  &   5 & 0.36 &  87 & 242 & 0.31 &  \cite{richey78} \\
  Cone Springs & kcal m$^{-2}$ yr$^{-1}$  &   5 & 0.32 & 11,819 & 30,626 & 0.09 &  \cite{tilly68} \\
  Silver Springs & kcal m$^{-2}$ yr$^{-1}$  &   5 & 0.28 & 21,296 & 29,175 & 0.00 &  \cite{odum57} \\
  English Channel & kcal m$^{-2}$ yr$^{-1}$  &   6 & 0.25 & 1,096 & 2,280 & 0.00 &  \cite{brylinsky72} \\
  Oyster Reef  & kcal m$^{-2}$ yr$^{-1}$  &   6 & 0.33 &  41 &  83 & 0.11 &  \cite{dame81} \\
  Baie de Somme & mgC m$^{-2}$ d$^{-1}$  &   9 & 0.30 & 876 & 2,034 & 0.14 &  \cite{rybarczyk03} \\
  Bothnian Bay & gC m$^{-2}$ yr$^{-1}$  &  12 & 0.22 &  44 & 183 & 0.23 &   \cite{sandberg00} \\
  Bothnian Sea & gC m$^{-2}$ yr$^{-1}$  &  12 & 0.24 & 117 & 562 & 0.31 &   \cite{sandberg00} \\
  Ythan Estuary & gC m$^{-2}$ yr$^{-1}$  &  13 & 0.23 & 1,258 & 4,181 & 0.24 &  \cite{baird81} \\
  Sundarban Mangrove (virgin) & kcal m$^{-2}$ yr$^{-1}$  &  14 & 0.22 & 117,959 & 441,214 & 0.16 & \cite{ray08} \\
  Sundarban Mangrove (reclaimed) & kcal m$^{-2}$ yr$^{-1}$  &  14 & 0.22 & 38,484 & 103,056 & 0.05 & \cite{ray08} \\
  Baltic Sea & mg C m$^{-2}$ d$^{-1}$  &  15 & 0.17 & 603 & 1,973 & 0.13 &   \cite{baird91} \\
  Ems Estuary & mg C m$^{-2}$ d$^{-1}$  &  15 & 0.19 & 282 & 1,067 & 0.32 &  \cite{baird91} \\
  Southern Benguela Upwelling & mg C m$^{-2}$ d$^{-1}$  &  16 & 0.23 & 714 & 2,545 & 0.31 & \cite{baird91} \\
  Peruvian Upwelling & mg C m$^{-2}$ d$^{-1}$  &  16 & 0.22 & 14,927 & 33,491 & 0.04 &  \cite{baird91} \\
  Crystal River (control) & mg C m$^{-2}$ d$^{-1}$  &  21 & 0.19 & 7,357 & 15,062 & 0.07 &  \cite{ulanowicz86} \\
  Crystal River (thermal) & mg C m$^{-2}$ d$^{-1}$  &  21 & 0.14 & 6,018 & 12,032 & 0.09 &  \cite{ulanowicz86} \\
  Charca de Maspalomas Lagoon & mg C m$^{-2}$ d$^{-1}$  &  21 & 0.12 & 1,486,230 & 6,010,331 & 0.18 &  \cite{almunia99} \\
  Northern Benguela Upwelling & mg C m$^{-2}$ d$^{-1}$  &  24 & 0.21 & 2,281 & 6,608 & 0.05 &  \cite{heymans00} \\
  Swartkops Estuary & mg C m$^{-2}$ d$^{-1}$  &  25 & 0.17 & 2,859 & 8,949 & 0.27 & \cite{scharler05} \\
  Sunday Estuary & mg C m$^{-2}$ d$^{-1}$  &  25 & 0.16 & 4,441 & 11,939 & 0.22 & \cite{scharler05} \\
  Kromme Estuary & mg C m$^{-2}$ d$^{-1}$  &  25 & 0.16 & 2,571 & 11,087 & 0.38 & \cite{scharler05} \\
  Okefenokee Swamp & g dw m$^{-2}$ y$^{-1}$  &  26 & 0.20 & 2,533 & 12,855 & 0.48 & \cite{whipple93} \\
  Neuse Estuary (early summer 1997) & mg C m$^{-2}$ d$^{-1}$  &  30 & 0.09 & 4,385 & 13,827 & 0.12 &  \cite{baird04} \\
  Neuse Estuary (late summer 1997)  & mg C m$^{-2}$ d$^{-1}$  &  30 & 0.11 & 4,639 & 13,035 & 0.13 &  \cite{baird04} \\
  Neuse Estuary (early summer 1998) & mg C m$^{-2}$ d$^{-1}$  &  30 & 0.09 & 4,568 & 14,025 & 0.12 &  \cite{baird04} \\
  Neuse Estuary (late summer 1998) & mg C m$^{-2}$ d$^{-1}$  &  30 & 0.10 & 5,641 & 15,031 & 0.11 &  \cite{baird04} \\
  Gulf of Maine & g ww m$^{-2}$ yr$^{-1}$  &  31 & 0.35 & 5,053 & 18,381 & 0.15 &   \cite{link08} \\
  Georges Bank & g ww m$^{-2}$ yr$^{-1}$  &  31 & 0.35 & 4,380 & 16,889 & 0.18 &  \cite{link08} \\
  Middle Atlantic Bight & g ww m$^{-2}$ yr$^{-1}$  &  32 & 0.37 & 4,869 & 17,916 & 0.18 &  \cite{link08} \\
  Narragansett Bay & mgC m$^{-2}$ yr$^{-1}$  &  32 & 0.15 & 693,845 & 3,917,246 & 0.51 &  \cite{monaco97} \\
  Southern New England Bight & g ww m$^{-2}$ yr$^{-1}$  &  33 & 0.35 & 4,717 & 17,597 & 0.16 &  \cite{link08} \\
  Chesapeake Bay  & mg C m$^{-2}$ yr$^{-1}$  &  36 & 0.09 & 888,791 & 3,227,453 & 0.19 &  \cite{baird89} \\
  Mondego Estuary (\emph{Zostera} sp. Meadows) & g AFDW m$^{-2}$ yr$^{-1}$ &  43 & 0.19 & 4,030 & 6,822 & 0.03 & \cite{patricio2004ascendency,patricio2006mass} \\
  St. Marks Seagrass, site 1 (Jan.) & mg C m$^{-2}$ d$^{-1}$  &  51 & 0.08 & 514 & 1,315 & 0.13 &  \cite{baird98} \\
  St. Marks Seagrass, site 1 (Feb.) & mg C m$^{-2}$ d$^{-1}$  &  51 & 0.08 & 601 & 1,590 & 0.11 &  \cite{baird98} \\
  St. Marks Seagrass, site 2 (Jan.) & mg C m$^{-2}$ d$^{-1}$  &  51 & 0.07 & 602 & 1,383 & 0.09 &  \cite{baird98} \\
  St. Marks Seagrass, site 2 (Feb.) & mg C m$^{-2}$ d$^{-1}$  &  51 & 0.08 & 800 & 1,921 & 0.08 &  \cite{baird98} \\
  St. Marks Seagrass, site 3 (Jan.) & mg C m$^{-2}$ d$^{-1}$  &  51 & 0.05 & 7,809 & 12,651 & 0.01 & \cite{baird98} \\
  St. Marks Seagrass, site 4 (Feb.) & mg C m$^{-2}$ d$^{-1}$  &  51 & 0.08 & 1,432 & 2,865 & 0.04 &  \cite{baird98} \\
  Sylt-R{\o}m{\o} Bight & mg C m$^{-2}$ d$^{-1}$  &  59 & 0.08 & 683,448 & 1,781,028 & 0.09 &  \cite{baird04_sylt} \\
  Graminoids (wet) & g C m$^{-2}$ yr$^{-1}$  &  66 & 0.18 & 6,272 & 13,676 & 0.02 &  \cite{ulanowicz00_graminoids} \\
  Graminoids (dry) & g C m$^{-2}$ yr$^{-1}$  &  66 & 0.18 & 3,472 & 7,519 & 0.04 &   \cite{ulanowicz00_graminoids} \\
  Cypress (wet) & g C m$^{-2}$ yr$^{-1}$  &  68 & 0.12 & 1,418 & 2,571 & 0.04 &  \cite{ulanowicz97_cypress} \\
  Cypress (dry) & g C m$^{-2}$ yr$^{-1}$  &  68 & 0.12 & 1,035 & 1,919 & 0.04 &  \cite{ulanowicz97_cypress} \\
  Lake Oneida (pre-ZM) & g C m$^{-2}$ yr$^{-1}$  &  74 & 0.22 & 1,034 & 1,697 & 0.00 &  \cite{miehls09_oneida} \\
  Lake Oneida (post-ZM) & g C m$^{-2}$ yr$^{-1}$  &  76 & 0.22 & 810 & 1,462 & 0.00 &  \cite{miehls09_oneida} \\
  Bay of Quinte (pre-ZM) & g C m$^{-2}$ yr$^{-1}$  &  74 & 0.21 & 989 & 1,517 & 0.00 &   \cite{miehls09_quinte} \\
  Bay of Quinte (post-ZM) & g C m$^{-2}$ yr$^{-1}$  &  80 & 0.21 & 1,163 & 2,107 & 0.01 &   \cite{miehls09_quinte} \\
  Mangroves (wet) & g C m$^{-2}$ yr$^{-1}$  &  94 & 0.15 & 1,531 & 3,265 & 0.10 &  \cite{ulanowicz99_mangrove} \\
  Mangroves (dry) & g C m$^{-2}$ yr$^{-1}$  &  94 & 0.15 & 1,531 & 3,272 & 0.10 &  \cite{ulanowicz99_mangrove} \\
  Florida Bay (wet) & mg C m$^{-2}$ yr$^{-1}$  & 125 & 0.12 & 738 & 2,720 & 0.14 &  \cite{ulanowicz98_fb} \\
  Florida Bay (dry) & mg C m$^{-2}$ yr$^{-1}$  & 125 & 0.13 & 547 & 1,778 & 0.08 &  \cite{ulanowicz98_fb} \\
   \hline
\end{tabular}
\end{tiny}
\end{center}
\end{table}



\end{document}